# Symmetry-Aware Predicate Abstraction for Shared-Variable Concurrent Programs (Extended Technical Report)

Alastair Donaldson, Alexander Kaiser, Daniel Kroening, and Thomas Wahl

Oxford University Computing Laboratory, United Kingdom

**Abstract.** *Predicate abstraction* is a key enabling technology for applying finite-state model checkers to programs written in mainstream languages. It has been used very successfully for debugging sequential system-level C code. Although model checking was originally designed for analyzing *concurrent* systems, there is little evidence of fruitful applications of predicate abstraction to shared-variable concurrent software. The goal of this paper is to close this gap. We have developed a *symmetry-aware* predicate abstraction strategy: it takes into account the replicated structure of C programs that consist of many threads executing the same procedure, and generates a Boolean program template whose multi-threaded execution soundly overapproximates the concurrent C program. State explosion during model checking parallel instantiations of this template can now be absorbed by exploiting symmetry. We have implemented our method in the SATABS predicate abstraction framework, and demonstrate its superior performance over alternative approaches on a large range of synchronization programs.

## 1 Introduction

*Concurrent software model checking* is one of the most challenging problems facing the verification community today. Not only does software generally suffer from *data state explosion*. Concurrent software in particular is susceptible to state explosion due to the need to track arbitrary thread interleavings, whose number grows exponentially with the number of executing threads.

*Predicate abstraction* [1] was introduced as a way of dealing with data state explosion: the program state is approximated via the values of a finite number of predicates over the program variables. Predicate abstraction turns C programs into finite-state *Boolean* programs [2], which can be model checked. Since insufficiently many predicates can cause spurious verification results, predicate abstraction is typically embedded into a counterexample-guided abstraction refinement (CEGAR) framework [3]. The feasibility of the overall approach has been convincingly demonstrated for *sequential* software by the success of the SLAM project at Microsoft, which was able to discover numerous control-dominated errors in low-level operating system code [4].

The majority of concurrent software is written using mainstream APIs such as POSIX threads (pthreads) in C/C++, or using a combination of language and library support, such as the Thread class, Runnable interface and synchronized construct in Java. Typically, multiple threads are spawned — up front or dynamically, in response to varying system load levels — to execute a given procedure in parallel, communicating via shared global variables. For such shared-variable concurrent programs, predicate abstraction success stories similar to that of SLAM are few and far between. The bottleneck is the exponential dependence of the generated state space on the number of running threads, which, if not addressed, permits exhaustive exploration of such programs only for trivial thread counts. The key to obtaining scalability is to exploit the *symmetry* naturally exhibited by these programs, namely their invariance under permutations of the participating threads.

Fortunately, much progress has recently been made on analyzing *replicated* Boolean programs, where a non-recursive Boolean template program is executed concurrently by many threads [5–7]. In this paper, we present an approach to predicate-abstracting concurrent programs that leverages this recent progress. More precisely:

– We describe a scheme to translate a non-recursive C program $\mathbb{P}$ with shared (global-scope) and local (procedure-scope) variables into a Boolean program $\mathbb{B}$ such that the $n$-thread Boolean program, denoted $\mathbb{B}^n$, soundly overapproximates the $n$-thread C program, denoted $\mathbb{P}^n$. We call such an abstraction method *symmetry-aware*.
– Our method permits predicates over arbitrary C program variables, local or global. We illustrate below the ramifications of this objective.

We also show in this paper how our approach can be implemented for C-like languages, complete with pointers and aliasing, and discuss the issues of spurious error detection and predicate refinement. In the rest of the Introduction, however, we illustrate why approaching the above two main goals naïvely can render abstraction **unsound**, creating the danger of missing bugs, which defies the purpose of reliable program verification.

A remark on notation: In program listings, we use == for the comparison operator, while = denotes assignment (as in C). Concurrent threads are assumed to interleave with statement-level granularity; see the discussion in the Conclusion on this subject.

### 1.1 Predicate Abstraction using Mixed Predicates

The Boolean program $\mathbb{B}$ to be built from the C program $\mathbb{P}$ will consist of Boolean variables, one per predicate as usual. Since $\mathbb{B}$ is to be executed by parallel threads, its variables have to be partitioned into "shared" and "local". As these variables track the values of various predicates over C program variables, the "shared" and "local" attributes clearly depend on the attributes of the C variables a predicate is formulated over. We therefore classify predicates as follows.

**Definition 1** *A **local** predicate refers solely to local C program variables. Analogously, a **shared** predicate refers solely to shared variables. A **mixed** predicate is neither local nor shared.*

We reasonably assume that each predicate refers to at least one program variable. A mixed predicate thus refers to both local and shared variables, and the above classification *partitions* the set of predicates into the three categories.

Given this classification, consider a local predicate $\phi$, which can change only as a result of a thread changing one of its local C variables; a change that is not visible to any other thread. This locality is inherited by the Boolean program if predicate $\phi$ is tracked by a local Boolean variable. Similarly, shared predicates are naturally tracked by shared Boolean variables.

For a mixed predicate, the decision whether it should be tracked in the shared or in the local space of the Boolean program is non-obvious. Consider first the following program $\mathbb{P}$ and the corresponding generated Boolean program $\mathbb{B}$, which tracks the mixed predicate $s \ != \ l$ in a **local** Boolean variable $b$:

$\mathbb{P}$:
```
0: shared int  s = 0;
   local  int  l = 1;
1: assert  s != l;
2: ++s;
```

$\mathbb{B}$:
```
0: local bool b = 1;
1: assert b;
2: b = b ? * : 1;
```

Consider the program $\mathbb{P}^2$, a two-thread instantiation of $\mathbb{P}$. It is easy to see that execution of $\mathbb{P}^2$ can lead to an assertion violation, while the corresponding concurrent Boolean program $\mathbb{B}^2$ is correct. (In fact, $\mathbb{B}^n$ is correct for any $n > 0$.) As a result, $\mathbb{B}^2$ is an **unsound** abstraction for $\mathbb{P}^2$. Consider now the following program $\mathbb{P}'$ and its Boolean abstraction $\mathbb{B}'$, which tracks the mixed predicate $s \ == \ l$ in a **shared** Boolean variable $b$:

$\mathbb{P}'$:
```
0: shared int   s = 0;
   shared bool  t = 1;
   local  int   l = 0;
1: assert t ↔ (s == l);
2: assume t;
3: ++l , t = 0;
```

$\mathbb{B}'$:
```
0: shared bool b = 1;
   shared bool t = 1;
1: assert t ↔ b;
2: assume t;
3: b , t = ( b ? 0 : * ) , 0;
```

Execution of $(\mathbb{P}')^2$ leads to an assertion violation if the first thread executes $\mathbb{P}'$ without interruption, while $(\mathbb{B}')^2$ is correct. We conclude that $(\mathbb{B}')^2$ is **unsound** for $(\mathbb{P}')^2$. The unsoundness can be eliminated by making $b$ local in $\mathbb{B}'$; an analogous reasoning removes the unsoundness in $\mathbb{B}$ as an abstraction for $\mathbb{P}$. It is clear from these examples, however, that in general a predicate of the form $s \ == \ l$ that genuinely depends on $s$ and $l$ cannot be tracked by a shared or a local variable without further amendments to the abstraction process.

At this point it may be useful to pause for a moment and consider whether, instead of designing solutions that deal with mixed predicates, we may not be better off by banning them, relying solely on shared and local predicates. Such restrictions on the choice of predicates render very simple bug-free programs **unverifiable** using predicate abstraction, including the following program $\mathbb{P}''$:

$\mathbb{P}''$:
```
0: shared int  r = 0;
   shared int  s = 0;
   local  int  l = 0;
1: ++r;
2: if (r == 1) then
3:    f();
```

$f()$:
```
4: ++s, ++l;
5: assert s == l;
6: goto 4;
```

The assertion in $\mathbb{P}''$ cannot be violated, no matter how many threads execute $\mathbb{P}$, since no thread but the first will manage to execute $f$. It is easy to prove that, over a set of **non-mixed** predicates (i.e. no predicate refers to both $l$ and one of $\{s, r\}$), no invariant is computable that is strong enough to prove $s == l$. We have included such a proof in Appendix B.

A technically simple solution to all these problems is to instantiate the template $\mathbb{P}$ $n$ times, once for each thread, into programs $\{\mathbb{P}_1, \ldots, \mathbb{P}_n\}$, in which indices $1, \ldots, n$ are attached to the local variables of the template, indicating the variable's owner. Every predicate that refers to local variables is similarly instantiated $n$ times. The new program has two features: (i) all its variables, having unambiguous names, can be declared at the global scope and are thus shared, including the original global program variables, and (ii) it is multi-threaded, but the threads no longer execute the same code. Feature (i) allows the new program to be predicate-abstracted in the conventional fashion: each predicate is stored in a shared Boolean variable. Feature (ii), however, entails that the new program is no longer symmetric. Model checking it will therefore have to bear the brunt of concurrency state explosion. Such an approach, which we refer to as *symmetry-oblivious*, will not scale beyond a very small number of threads.

To summarize our findings: Mixed predicates are necessary to prove properties for even very simple programs. On the other hand, they cannot be tracked using standard thread-local or shared variables. Disambiguating local variables avoids mixed predicates, but destroys symmetry. It is the goal of this paper to design a solution without the need to compromise.

## 2 Symmetry-Aware Predicate Abstraction

In order to illustrate our method, let $\mathbb{P}$ be a program defined over a set of variables $V$ that is partitioned in the form $V = V_S \cup V_L$ into *shared* and *local* variables. The parallel execution of $\mathbb{P}$ by $n$ threads is a program defined over the shared variables and $n$ copies of the local variables, one copy for each thread. A thread is nondeterministically chosen to be *active*, i.e. to execute a statement of $\mathbb{P}$, potentially modifying the shared variables, and its own local variables, but nothing else. In this section, we ignore the specific syntax of statements, and we do not consider language features that introduce aliasing, such as pointers (these are the subject of Section 3). Therefore, an assignment to a variable $v$ cannot modify a variable other than $v$, and an expression $\phi$ depends only on the variables occurring in it, which we refer to as $Loc(\phi) = \{v : v \text{ occurs in } \phi\}$.

### 2.1 Mixed Predicates and Notify-All Updates

Our goal is to translate the program $\mathbb{P}$ into a Boolean program $\mathbb{B}$ such that, for any $n$, a suitably defined parallel execution of $\mathbb{B}$ by $n$ threads overapproximates the parallel execution of $\mathbb{P}$ by $n$ threads. Let $E = \{\phi_1, \ldots, \phi_m\}$ be a set of predicates over $\mathbb{P}$, i.e. a set of Boolean expressions over variables in $V$. We say $\phi_i$ is

**shared** if $Loc(\phi_i) \subseteq V_S$,
 **local** if $Loc(\phi_i) \subseteq V_L$, and
**mixed** otherwise, i.e. $Loc(\phi_i) \cap V_L \neq \emptyset$ and $Loc(\phi_i) \cap V_S \neq \emptyset$.

We declare, in $\mathbb{B}$, Boolean variables $\{b_1, \ldots, b_m\}$; the intention is that $b_i$ tracks the value of $\phi_i$ during abstract execution of $\mathbb{P}$. We partition these Boolean variables into *shared* and *local* by stipulating that $b_i$ is shared if $\phi_i$ is shared; otherwise $b_i$ is local. In particular, **mixed predicates are tracked in local variables**. Intuitively, the value of a mixed predicate $\phi_i$ depends on the thread it is evaluated over. Declaring $b_i$ shared would thus necessarily lose information. Declaring it local does not lose information, but, as the example in the Introduction has shown, is insufficient to guarantee a sound abstraction. We will see shortly how to solve this problem.

Each statement in $\mathbb{P}$ is now translated into a corresponding statement in $\mathbb{B}$. Statements related to flow of control are handled using techniques from standard predicate abstraction [2]; the distinction between shared, mixed and local predicates does not matter here. Consider an assignment to a variable $v$ in $\mathbb{P}$ and a Boolean variable $b$ of $\mathbb{B}$ with associated predicate $\phi$. We first check whether variable $v$ affects $\phi$, written $\mathit{affects}(v, \phi)$. Given that in this section we assume no aliasing, this is the case exactly if $v \in \mathit{Loc}(\phi)$. If $\mathit{affects}(v, \phi)$ evaluates to *false*, $b$ does not change. Otherwise, code needs to be generated to update $b$. This code needs to take into account the "flavors" of $v$ and $\phi$, which give rise to three different flavors of updates of $b$:

**shared update:** Suppose $v$ and $\phi$ are both shared. An assignment to $v$ is visible to all threads, so the truth of $\phi$ is modified for all threads. This is reflected in $\mathbb{B}$: by our stipulation above, the shared predicate $\phi$ is tracked by the *shared* variable $b$. Thus, we simply generate code to update $b$ according to standard sequential predicate abstraction rules; the new value of $b$ is shared among all threads.

**local update:** Suppose $v$ is local and $\phi$ is local or mixed. An assignment to $v$ is visible only by the active (executing) thread, so the truth of $\phi$ is modified only for the active thread. This also is reflected in $\mathbb{B}$: by our stipulation above, the local or mixed predicate $\phi$ is tracked by the *local* variable $b$. Again, sequential predicate abstraction rules suffice; the value of $b$ changes only for the active thread.

**notify-all update:** Suppose $v$ is shared and $\phi$ is mixed. An assignment to $v$ is visible to all threads, so the truth of $\phi$ is modified for all threads. This is **not** reflected in $\mathbb{B}$: by our stipulation above, the mixed predicate $\phi$ is tracked by the *local* variable $b$, which will be updated only by the active thread. We will solve this problem by (i) generating code to update $b$ according to standard sequential predicate abstraction rules, and (ii) **notifying** all *passive* threads of the modification of the shared variable $v$, so as to allow them to update their local copy of $b$.

We write $\mathit{must\_notify}(v, \phi)$ if the shared variable $v$ affects the mixed predicate $\phi$:

$$\mathit{must\_notify}(v, \phi) \quad = \quad \mathit{affects}(v, \phi) \,\wedge\, v \in V_S \,\wedge\, (\mathit{Loc}(\phi) \cap V_L \neq \emptyset)\,.$$

This formula evaluates to *true* exactly when it is necessary to notify passive threads of an update to $v$. What remains to be discussed in the rest of this section is how notifications are implemented in $\mathbb{B}$.

### 2.2 Implementing Notify-All Updates

We pause to recap some terminology and notation from sequential predicate abstraction [2]. Given a set $E = \{\phi_1, \ldots, \phi_m\}$ of predicates tracked by variables $\{b_1, \ldots, b_m\}$, an assignment statement $st$ is translated into the following code, in parallel for each $i \in \{1, \ldots, m\}$:

$$
\begin{aligned}
&\textbf{if} && \mathcal{F}(\mathit{WP}(\phi_i, st)) \textbf{ then } b_i &&= 1 \\
&\textbf{else if } \mathcal{F}(\mathit{WP}(\neg\phi_i, st)) \textbf{ then } b_i &&= 0 \\
&\textbf{else} && b_i &&= \star.
\end{aligned} \quad (1)
$$

Here, $\star$ is the nondeterministic choice expression, $\mathit{WP}$ the weakest precondition operator, and $\mathcal{F}$ the operator that strengthens an arbitrary predicate to a disjunction of cubes over the $b_i$. For example, with predicate $\phi :: (l < 10)$ tracked by variable $b$, $E = \{\phi\}$ and statement $st :: \mathtt{++}l$, we obtain $\mathcal{F}(\mathit{WP}(\phi, st)) = \mathcal{F}(l < 9) = \mathit{false}$ and $\mathcal{F}(\mathit{WP}(\neg\phi, st)) = \mathcal{F}(l \mathtt{>=} 9) = (l \mathtt{>=} 10) = \neg\phi$, so that (1) reduces to

$$b = ( b \ ? \ \star \ : \ 0 ).$$

In general, (1) is often abbreviated using the assignment

$$b_i = \mathit{choose}(\mathcal{F}(\mathit{WP}(\phi_i, st)), \mathcal{F}(\mathit{WP}(\neg\phi_i, st))),$$

where $\mathit{choose}(x, y)$ returns 1 if $x$ evaluates to $\mathit{true}$, 0 if $(\neg x) \wedge y$ evaluates to $\mathit{true}$, and $\star$ otherwise. Abstraction of control flow guards uses the $\mathcal{G}$ operator, which is dual to $\mathcal{F}$: $\mathcal{G}(\phi) = \neg\mathcal{F}(\neg(\phi))$.

Returning to symmetry-aware predicate abstraction, if $\mathit{must\_notify}(v, \phi)$ evaluates to $\mathit{true}$ for $\phi$ and $v$, predicate $\phi$ is mixed and thus tracked in $\mathbb{B}$ by some local Boolean variable, say $b$. Predicate-abstracting an assignment of the form $v = \chi$ requires updating the active thread's copy of $b$, as well as broadcasting an instruction to all passive threads to update their copy of $b$, in view of the new value of $v$. This is implemented using two assignments, which are executed in parallel. The first assignment is as follows:

$$b = \mathit{choose}(\mathcal{F}(\mathit{WP}(\phi, v = \chi)), \mathcal{F}(\mathit{WP}(\neg\phi, v = \chi))). \quad (2)$$

This assignment has standard predicate abstraction semantics:[1] variable $b$ of the active thread is updated by computing the weakest precondition of predicate $\phi$ and its negation with respect to the statement $v = \chi$, applying the strengthening operator $\mathcal{F}$ to make the precondition expressible over the existing predicates, and by applying the $\mathit{choose}$ operator, which may introduce nondeterminism. Note that, since expression $\chi$ involves only local variables of the active thread and shared variables, only predicates over those variables are involved in the defining expression for $b$.

The second assignment looks similar, but introduces a new symbol:

$$[b] = \mathit{choose}(\mathcal{F}(\mathit{WP}([\phi], v = \chi)), \mathcal{F}(\mathit{WP}(\neg[\phi], v = \chi))). \quad (3)$$

The notation $[b]$ stands for the copy of local variable $b$ owned by some passive thread. Similarly, $[\phi]$ stands for the expression defining predicate $\phi$, but with every local variable occurring in the expression replaced by the copy owned by the passive thread; this is the predicate $\phi$ in the context of the passive thread. Weakest precondition computation is with respect to $[\phi]$, while the assignment $v = \chi$, as an argument to $\mathit{WP}$, is unchanged: $v$ is shared, and local variables appearing in the defining expression $\chi$ must be interpreted as local variables of the *active* thread. Assignment (3) has the effect of updating variable $b$ in every passive thread. We call Boolean programs involving assignments of the form $[b] = \ldots$ *Boolean broadcast programs*; a formal syntax and semantics for such programs is given in Appendix A.

---

[1] Our presentation is in terms of the Cartesian abstraction [8], as used in [2], but our method in general is independent of the abstraction used.

Let us illustrate the above technique using a canonical example: consider the assignment $s = l$, for shared and local variables $s$ and $l$, and define the mixed predicate $\phi :: (s == l)$. The first part of the above parallel assignment simplifies to $b = true$. For the second part, we obtain:

$$[b] = choose(\mathcal{F}(WP(s==[l]), s=l)), \mathcal{F}(WP(\neg(s==[l]), s=l))) \,.$$

Computing weakest preconditions, this reduces to:

$$[b] = choose(\mathcal{F}(l==[l]), \mathcal{F}(\neg(l==[l]))) \,.$$

*Precision of the Abstraction.* To evaluate this expression further, we have to decide on the set of predicates available to the $\mathcal{F}$ operator to express the preconditions. If this set includes only predicates over the shared variables and the local variables of the passive thread that owns $[b]$, the predicate $l == [l]$ is not expressible and must be strengthened to $false$. The above assignment then simplifies to $[b] = choose(false, false)$, i.e. $[b] = \star$. The mixed predicates owned by passive threads are essentially *invalidated* when the active thread modifies a shared variable occurring in such predicates, resulting in a very imprecise abstraction.

We can exploit information stored in predicates local to other threads, to increase the precision of the abstraction. For maximum precision one could make *all* other threads' predicates available to the strengthening operator $\mathcal{F}$. This happens in the symmetry-oblivious approach sketched in the Introduction, where local and mixed predicates are physically replicated and declared at the global scope and can thus be made available to $\mathcal{F}$. Not surprisingly, in practice, replicating predicates in this way renders the abstraction prohibitively expensive. We analyze this experimentally in Section 5.

A compromise which we have found to work well in practice (again, demonstrated in Section 5) is to equip operator $\mathcal{F}$ with all shared predicates, all predicates of the passive thread owning $[b]$, **and also** predicates of the active thread. This arrangement is intuitive since the update of a passive thread's local variable $[b]$ is due to an assignment performed by some active thread. Applying this compromise to our canonical example: if both $s == [l]$ and $s == l$ evaluate to true before the assignment $s=l$, we can conclude that $[l] == l$ before the assignment, and hence $s == [l]$ after the assignment. Using $\oplus$ to denote exclusive-or, the assignment to $[b]$ becomes:

$$[b] = choose([b] \wedge b, [b] \oplus b) \,.$$

### 2.3 The Predicate Abstraction Algorithm

We now show how our technique for soundly handling mixed predicates is used in an algorithm for predicate abstracting C-like programs. To present the algorithm compactly, we assume a language with three types of statement: assignments, nondeterministic gotos, and assumptions. Control-flow can be modelled via a combination of gotos and assumes, in the standard way.

**Algorithm 1** Predicate abstraction

**Input**: Program template $\mathbb{P}$, set of predicates $\{\phi_1, \ldots, \phi_m\}$
**Output**: Boolean program $\mathbb{B}$ over variables $b_1, \ldots, b_m$

1: **for each** statement $d$: stmt of $\mathbb{P}$ **do**
2:    **if** stmt is $v = \psi$ **then**
3:       $\{i_1, \ldots, i_f\} \leftarrow \{i \mid 1 \leq i \leq m \land \mathit{affects}(v, \phi_i)\}$
4:       $\{j_1, \ldots, j_g\} \leftarrow \{j \mid 1 \leq i \leq m \land \mathit{must\_notify}(v, \phi_i)\}$
5:       output $d$:
$$\begin{pmatrix} b_{i_1} \\ \vdots \\ b_{i_f} \end{pmatrix} = \begin{pmatrix} choose(\mathcal{F}(WP(\phi_{i_1}, v{=}\psi)), \mathcal{F}(WP(\neg\phi_{i_1}, v{=}\psi))) \\ \vdots \\ choose(\mathcal{F}(WP(\phi_{i_f}, v{=}\psi)), \mathcal{F}(WP(\neg\phi_{i_f}, v{=}\psi))) \end{pmatrix},$$
$$\begin{pmatrix} [b_{j_1}] \\ \vdots \\ [b_{j_g}] \end{pmatrix} = \begin{pmatrix} choose(\mathcal{F}(WP([\phi_{j_1}], v{=}\psi)), \mathcal{F}(WP(\neg[\phi_{j_1}], v{=}\psi))) \\ \vdots \\ choose(\mathcal{F}(WP([\phi_{j_g}], v{=}\psi)), \mathcal{F}(WP(\neg[\phi_{j_g}], v{=}\psi))) \end{pmatrix}$$
6:    **else if** stmt is **goto** $d_1, \ldots, d_m$ **then**
7:       output $d$: **goto** $d_1, \ldots, d_m$;
8:    **else if** stmt is **assume** $\phi$ **then**
9:       output $d$: **assume** $\mathcal{G}(\phi)$;

Algorithm 1 processes an input program template of this form and outputs a corresponding Boolean broadcast program template. Statements **goto** and **assume** are handled as in standard predicate abstraction: the former are left unchanged, while the latter are translated directly except that the guard of an **assume** statement is expressed over Boolean program variables using the $\mathcal{G}$ operator (see Section 2.2).

The interesting part of the algorithm for us is the translation of assignment statements. For each assignment, a corresponding parallel assignment to Boolean program variables is generated. The *affects* and *must_notify* predicates are used to decide for which Boolean variables regular and broadcast assignments are required, respectively.

## 3 Symmetry-Aware Predicate Abstraction with Aliasing

In Section 2 we outlined our novel predicate abstraction technique, ignoring complications introduced by pointers and aliasing. We now explain how symmetry-aware predicate abstraction is realized in practice, for C programs that manipulate pointers. We impose one restriction: we do not consider programs where a shared pointer variable, or a pointer variable local to thread $i$, can point to a variable local to thread $j$ (with $j \neq i$). This scenario arises only when a thread copies the address of a stack or thread-local variable to the shared state. This unusual programming style allows thread $i$ to directly modify the local state of thread $j$ at the C program level, breaking the asynchronous model of computation assumed by our method.

For ease of presentation we consider the scenario where program variables either have a base type (e.g. **int** or **float**), or pointer type (e.g. **int**$\star$ or **float**$\star\star$). Our method can be extended to handle records, arrays and heap-allocated memory in a straightfor-ward but laborious manner. As in [2], we also assume that input programs have been processed so that l-values involve at most one pointer dereference.

Alias information is important in deciding, once and for all, whether predicates should be classed as local, mixed or shared. For example, let $p$ be a local variable of type **int**$\star$, and consider predicate $\phi :: (\star p == 1)$. Clearly $\phi$ is not shared since it depends on local variable $p$. Whether $\phi$ should be regarded as a local or mixed predicate depends on whether $p$ may point to the shared state: we regard $\phi$ as local if $p$ can never point to a shared variable, otherwise $\phi$ is classed as mixed. Alias information also lets us determine whether a variable update may affect the truth of a given predicate, and whether it is necessary to notify other threads of this update. We now show how these intuitions can be formally integrated with our predicate abstraction technique. This involves suitably refining the notions of local, shared and mixed predicates, and the definitions of $\mathit{affects}$ and $\mathit{must\_notify}$ introduced in Section 2 and used by Algorithm 1.

### 3.1 Aliasing, Locations of Expressions, and Targets of l-values

We assume the existence of a sound pointer alias analysis for concurrent programs, e.g. [9], which we treat as a black box. This procedure conservatively tells us whether a shared variable with pointer type may refer to a local variable. As discussed at the start of Section 3, we reject programs where this is the case.[2] Otherwise, for a program template $\mathbb{P}$ over variables $V$, alias analysis yields a relation $\mapsto_d \subseteq V \times V$ for each program location $d$. For $v, w \in V$, if $v \not\mapsto_d w$ then $v$ provably does not point to $w$ at $d$.

For an expression $\phi$ and program point $d$, we write $loc(\phi, d)$ for the set of variables that it may be necessary to access in order to evaluate $\phi$ at $d$, during an arbitrary program run. The precision of $loc(\phi, d)$ is directly related to the precision of alias analysis.

**Definition 2** *For a constant value $z$, $v \in V$ and $k > 0$, we define:*

$$loc(z, d) = \emptyset \quad loc(\&v, d) = \emptyset \quad loc(v, d) = \{v\}$$

$$loc(\underbrace{\star \ldots \star}_{k} v, d) = \{v\} \cup \bigcup_{w \in V} \{loc(\underbrace{\star \ldots \star}_{k-1} w, d) \mid v \mapsto_d w\}$$

*For other compound expressions, $loc(\phi, d)$ is defined recursively in the obvious way.*

Definition 2 captures the fact that evaluating a pointer dereference $\star v$ involves reading both $v$ and the variable to which $v$ points, while evaluating an "address-of" expression, $\&v$, requires no variable accesses: addresses of variables are fixed at compile time.

**Definition 3** *For an expression $\phi$, $Loc(\phi)$ is the set of variables that may need to be accessed to evaluate $\phi$ at an* arbitrary *program point during an arbitrary program run:*

$$Loc(\phi) = \bigcup_{1 \leq d \leq k} loc(\phi, d).$$

Note how this definition of $Loc$ generalizes that used in Section 2.

---

[2] This also eliminates the possibility of thread $i$ pointing to variables in thread $j \neq i$: the address of a variable in thread $j$ would have to be communicated to thread $i$ via a shared variable.

**Definition 4** *We write $targets(x, d)$ for the set of variables that may be modified by writing to l-value $x$ at program point d:*

$$targets(v, d) = \{v\} \qquad targets(\star v, d) = \{w \in V \mid v \mapsto_d w\}$$

Note that we have $targets(\star v, d) \neq loc(\star v, d)$. This is because *writing* through $\star v$ modifies only the variable to which $v$ points, while *reading* the value of $\star v$ involves reading the value of $v$, to determine which variable $w$ is pointed to by $v$, and then reading the value of $w$.

### 3.2 Shared, Local and Mixed Predicates in the Presence of Aliasing

In the presence of pointers, we define the notion of a predicate $\phi$ being shared, local, or mixed exactly as in Section 2.1, only with the generalization of *Loc* presented in Definition 3. In Section 2.1, without pointers, we could classify $\phi$ purely syntactically, based on whether any shared variables appear in $\phi$. In the presence of pointers, we must classify $\phi$ with respect to alias information; our definition of *Loc* takes care of this.

Recall from Section 2.1 that we defined $\mathit{affects}(v, \phi) = (v \in Loc(\phi))$ to indicate that updating variable $v$ may affect the truth of predicate $\phi$. In the presence of pointers, this definition no longer suffices. The truth of $\phi$ may be affected by assigning to l-value $x$ if $x$ may alias some variable on which $\phi$ depends. Whether this is the case depends on the program point at which the update occurs. Definitions 2 and 4 of *loc* and *targets* allow us to express this:

$$\mathit{affects}(x, \phi, d) = (targets(x, d) \cap loc(\phi, d) \neq \emptyset).$$

We also need to determine whether an update affects the truth of a predicate only for the thread executing the update, or for all threads. The definition of $\mathit{must\_notify}$ presented in Section 2.1 needs to be adapted to take aliasing into account. At first sight, it seems that we must simply parameterise *affects* according to program location, and replace the conjunct $v \in V_S$ with the condition that $x$ may target some shared variable:

$$\mathit{must\_notify}(x, \phi, d) = \mathit{affects}(x, \phi, d) \land (Loc(\phi) \cap V_L \neq \emptyset) \\ \land (targets(x, d) \cap V_S \neq \emptyset).$$

However, this is unnecessarily strict. We can refine the above definition to minimise the extent to which notifications are required, as follows:

$$\mathit{must\_notify}(x, \phi, d) = (targets(x, d) \cap Loc(\phi) \cap V_S \neq \emptyset) \land (Loc(\phi) \cap V_L \neq \emptyset).$$

The refined definition avoids the need for thread notification in the following scenario. Suppose we have shared variables $s$ and $t$, local variable $l$, local pointer variable $p$, and predicate $\phi :: (s > l)$. Consider an assignment to $\star p$ at program point $d$. Suppose that alias analysis tells us exactly $p \mapsto_d t$ and $p \mapsto_d l$. The only shared variable that can be modified by assigning through $\star p$ at program point $d$ is $t$, and the truth of $\phi$ does not depend on $t$. Thus the assignment does *not* require a "notify-all" with respect to $\phi$. Working through the definitions, we find that our refinement of $\mathit{must\_notify}$ correctly determines this, while the direct extension of *must_notify* from Section 2.1 would lead to an unnecessary "notify-all".

The predicate abstraction algorithm, Algorithm 1, can now be adapted to handle pointers: parameter $d$ is simply added to the uses of *affects* and *must_notify*. Handling of pointers in weakest preconditions works as in standard predicate abstraction [2], using Morris's general axiom of assignment [10].

## 4  Closing the CEGAR Loop

So far we have presented a novel technique for predicate-abstracting symmetric concurrent programs. We have integrated our method with the SATABS CEGAR-based verifier [11], using the Cartesian abstraction method [8] and the maximum cube length approximation [2]. We now sketch how we have adapted the other phases of the CEGAR loop: model checking, simulation and refinement, to accurately handle concurrency; a detailed description of the entire process is left to an extended version of this paper.

**Model checking Boolean broadcast programs.** Our predicate abstraction technique generates a concurrent Boolean broadcast program. The extended syntax and semantics for broadcasts mean that we cannot simply use existing concurrent Boolean program model checkers, such as BOOM [12] or BOPPO [13], for the model checking phase of the CEGAR loop. We have implemented a prototype extension of BOOM, which we call B-BOOM. B-BOOM extends the counter abstraction-based symmetry reduction capabilities of BOOM [5] to support broadcast operations. Symbolic image computation for broadcast assignments is significantly more expensive than image computation for standard assignments. In the context of BOOM it involves 1) converting states from counter representation to a form where the individual local states of threads are stored using distinct BDD variables, 2) computing the intersection of $n-1$ successor states, one for each inactive thread paired with the active thread, and 3) transforming the resulting state representation back to counter form using Shannon expansion. The expense of image computation for broadcasts motivates the careful analysis we have presented in Sections 2 and 3 for determining tight conditions under which broadcasts are required.

**Simulation.** To determine the authenticity of abstract error traces reported by B-BOOM we have extended the SATABS simulator. The existing simulator extracts the control flow from the trace. This is mapped back to the original C program and translated into a propositional formula (using standard techniques such as single static assignment conversion and bitvector interpretation of variables). The error is spurious exactly if this formula is unsatisfiable. In the concurrent case, the control flow information of an abstract trace includes which thread executes actively in each step. We have extended the simulator so that each local variable involved in a step is replaced by a fresh indexed version, indicating the executing thread that owns the variable. The result is a program trace over the replicated C program $\mathbb{P}^n$, which can be directly checked using a SAT solver.

As an example, suppose function $f$ from program $\mathbb{P}''$ (Introduction) is executed by 2 threads (for this example, we ignore the rest of $\mathbb{P}''$). The model checker may return

the error trace shown below on the left, which is converted into the non-threaded form
shown on the right.

```
T1: ++s, ++l;
T1: assert s == l;
T2: ++s, ++l;
T2: assert s == l;
```

```
++s, ++l_1;
assert s == l_1;
++s, ++l_2;
assert s == l_2;
```

The trace on the right is translated into SSA form and finally into the integer arithmetic
formula below, which is shown to be satisfiable, so the error is real.

$$s^0 = 0 \land l_1^0 = 0 \land l_2^0 = 0 \land s^1 = s^0 + 1 \land l_1^1 = l_1^0 + 1$$
$$\land s^2 = s^1 + 1 \land l_2^1 = l_2^0 + 1 \land s^2 \neq l_2^1,$$

Note that broadcast operations do not affect simulation: although, at the Boolean program level, a broadcast may change the state of multiple threads simultaneously, the corresponding C program statement is simply an update to a shared variable executed by a single thread.

**Refinement.** Our implementation performs refinement by extracting new predicates from counterexamples via weakest precondition calculations. This standard method requires a small modification in our context: weakest precondition calculations generate predicates over shared variables, and local variables of *specific* threads. For example, if thread 1 branches according to a condition such as $l < s$, where $l$ and $s$ are local and shared, respectively, weakest precondition calculations generate the predicate $l_1 < s$, where $l_1$ is thread 1's copy of $l$. Because our predicate abstraction technique works at the template program level, we cannot add this predicate directly. Instead, we generalize such predicates by removing thread indices. Hence in the above example, we add the mixed predicate $l < s$, for *all* threads.

An alternative approach is to refine the abstract transition relation associated with the Cartesian abstraction based on infeasible steps in the abstract counterexample [14]. We do not yet perform such refinement, due to challenge of correctly refining abstract transitions involving broadcast assignments. This involves some subtle issues which will require further research to solve.

## 5 Experimental Results

We evaluate the SATABS-based implementation of our techniques using a set of 14 concurrent C programs. We consider benchmarks where threads synchronize via locks (lock-based), or in a lock-free manner via atomic *compare-and-swap* (cas) or *test-and-set* (tas) instructions. The benchmarks are as follows:[3]

– **Increment, Inc./Dec. (lock-based and cas-based)** A counter, concurrently incremented, or incremented and decremented, by multiple threads [15]
– **Prng (lock-based and cas-based)** Concurrent pseudorandom number generator [15]

---

[3] All benchmarks and tools are available online: http://www.cprover.org/SAPA

- **Stack (lock-based and cas-based)** Thread-safe stack implementation, supporting concurrent pushes and pops, adapted from an Open Source IBM implementation[4] of an algorithm described in [15]
- **Tas Lock**, **Ticket Lock (tas-based)** Concurrent lock implementations [16]
- **FindMax**, **FindMaxOpt (lock-based and cas-based)** Implementations of parallel reduction operation [17] to find maximum element in array. **FindMax** is a basic implementation, and **FindMaxOpt** and optimized version where threads reduce communication by computing a partial maximum value locally

Mixed predicates were required for verification to succeed in all but two benchmarks: lock-based *Prng*, and lock-based *Stack*. For each benchmark, we consider verification of a simple safety property, specified via an assertion. We have also prepared a buggy version of each benchmark, where an error is injected into the source code to make it possible for this assertion to fail. We refer to correct and buggy versions of our benchmarks as *safe* and *unsafe*, respectively.

All experiments are performed on a 3GHz Intel Xeon machine with 40 GB RAM, running 64-bit Linux, with separate timeouts of 1h for the abstraction and model checking phases of the CEGAR loop. Predicate abstraction uses a maximum cube length of 3 for all examples, and MiniSat 2 (compiled with full optimizations) is used for predicate abstraction and counterexample simulation.

| Benchmark | $n$ | Symmetry Oblivious #$\phi$ | Abs (s) | MC (s) | Symmetry Aware #$\phi$ | Abs (s) | MC (s) | Benchmark | $n$ | Symmetry Oblivious #$\phi$ | Abs (s) | MC (s) | Symmetry Aware #$\phi$ | Abs (s) | MC (s) |
|---|---|---|---|---|---|---|---|---|---|---|---|---|---|---|---|
| Increment | 6 | 14 | 13 | 5 | 4 | **1** | **<1** | Prng | 6 | 57 | 69 | 21 | 13 | **26** | **<1** |
| (lock-based) | 8 | 18 | 29 | 152 | 4 | **1** | **1** | (lock-based) | 7 | 68 | 83 | 191 | 13 | **26** | **1** |
| | 9 | 20 | 40 | 789 | 4 | **1** | **1** | | 8 | 79 | 96 | T.O. | 13 | **26** | **2** |
| | 10 | 22 | 56 | T.O. | 4 | **1** | **2** | | 16 | – | – | – | 13 | **26** | **142** |
| | 12 | – | – | – | 4 | **1** | **7** | | 26 | – | – | – | 13 | **26** | **3023** |
| | 14 | – | – | – | 4 | **1** | **24** | Prng | 3 | 32 | **29** | <1 | 17 | 48 | **1** |
| | 16 | – | – | – | 4 | **1** | **100** | (cas-based) | 4 | 47 | 40 | 12 | 17 | 48 | **38** |
| | 18 | – | – | – | 4 | **1** | **559** | | 5 | 62 | 57 | 1049 | 17 | **48** | **1832** |
| | 20 | – | – | – | 4 | **1** | **2882** | FindMax | 6 | 6 | 5 | 30 | 1 | **<1** | **<1** |
| Increment | 4 | 26 | 201 | 12 | 8 | **6** | **1** | (lock-based) | 7 | 7 | 9 | 244 | 1 | **<1** | **1** |
| (cas-based) | 5 | 32 | 94 | 358 | 8 | **6** | **13** | | 8 | 8 | 14 | T.O. | 1 | **<1** | **1** |
| | 6 | 38 | 190 | T.O. | 8 | **6** | **116** | | 16 | – | – | – | 1 | **<1** | **125** |
| | 7 | – | – | – | 8 | **6** | **997** | | 25 | – | – | – | 1 | **<1** | **3005** |
| Inc./Dec. | 4 | 26 | 71 | 6 | 11 | **11** | **2** | FindMax | 3 | 18 | 4 | 7 | 6 | **1** | **2** |
| (lock-based) | 5 | 31 | 132 | 656 | 11 | **11** | **50** | (cas-based) | 4 | 24 | 8 | 407 | 6 | **1** | **368** |
| | 6 | 36 | 231 | T.O. | 11 | **11** | **1422** | FindMaxOpt | 4 | 8 | 3 | 40 | 2 | **<1** | **3** |
| Inc./Dec. | 3 | 45 | 372 | 6 | 20 | **78** | **3** | (lock-based) | 5 | 10 | 6 | 1356 | 2 | **<1** | **33** |
| (cas-based) | 4 | 58 | 872 | 4043 | 20 | **78** | **252** | | 6 | 12 | 11 | T.O. | 2 | **<1** | **269** |
| Tas Lock | 4 | 28 | 9 | 114 | 8 | **1** | **4** | | 7 | – | – | – | 2 | **<1** | **1773** |
| (tas-based) | 5 | 34 | 14 | T.O. | 8 | **1** | **72** | FindMaxOpt | 3 | 21 | 9 | 11 | 7 | **3** | **2** |
| | 6 | – | – | – | 8 | **1** | **725** | (cas-based) | 4 | 28 | 15 | 1097 | 7 | **3** | **61** |
| Ticket Lock | 2 | 22 | 554 | 1 | 19 | **251** | **1** | | 5 | 35 | 22 | T.O. | 7 | **3** | **1240** |
| (tas-based) | 3 | 27 | 1319 | 3 | 19 | **251** | **1** | Stack | 3 | 16 | <1 | 14 | 5 | **<1** | **8** |
| | 4 | 32 | T.O. | – | 19 | **251** | **2** | (lock-based) | 4 | 17 | <1 | 945 | 5 | **<1** | **374** |
| | 6 | – | – | – | 19 | **251** | **62** | Stack | 3 | 16 | 2 | 29 | 6 | **<1** | **14** |
| | 8 | – | – | – | 19 | **251** | **2839** | (cas-based) | 4 | 21 | 8 | 3408 | 6 | **<1** | **813** |

**Table 1.** Comparison of symmetry-aware and symmetry-oblivious predicate abstraction over our benchmarks. For each configuration, the fastest abstraction and model checking times are in bold.

---
[4] http://amino-cbbs.sourceforge.net

**Symmetry-aware vs. symmetry-oblivious method.** We evaluate the scalability of our symmetry-aware predicate abstraction technique (SAPA) by comparing it against the *symmetry-oblivious* predicate abstraction (SOPA) approach described in Section 1, for verification of correct versions of our benchmarks. Recall that in SOPA, an $n$-thread symmetric concurrent program is expanded so that variables for all threads are explicitly duplicated, and $n$ copies of all non-shared predicate are generated. The expanded program is then abstracted over the expanded set of predicates, using standard predicate abstraction. This yields a Boolean program for each thread; the parallel composition of these $n$ Boolean programs is explored by a model checker. Because symmetry is not exploited, and no broadcasts are required, any suitable model checker can be used. We have tried both standard BOOM [5] (without symmetry reduction) and Cadence SMV [18] to model check expanded Boolean programs. In all cases, we found BOOM to be faster than SMV, thus we present results only for BOOM.

Table 1 presents the results of the comparison. For each benchmark and each approach we show, for interesting thread counts (including the largest thread count that could be verified with each approach), the number of predicates required for verification and the elapsed time for predicate abstraction and model checking. For each configuration, the fastest abstraction and model checking times are shown in bold. Model checking uses standard BOOM, without symmetry reduction (SOPA) and B-BOOM, our novel extension to BOOM discussed in Section 4 (SAPA), respectively. Entries marked T.O. indicate that a timeout occurred; succeeding cells are then marked '–'.

The results show that our novel SAPA technique significantly outperforms SOPA, both in terms of abstraction and model checking time. The former can be attributed to the fact that, with SOPA, the number of predicates grows according to the number of threads considered, while with SAPA, this is thread count-independent. The latter is due to the exploitation of template-level symmetry by B-BOOM.

| Benchmark | Symmetry-Aware | | Mixed as local | | Mixed as shared | |
|---|---|---|---|---|---|---|
| | Safe $n$ | Unsafe $n$ | Safe $n$ | Unsafe $n$ | Safe $n$ | Unsafe $n$ |
| Increment (lock-based) | safe >10 | unsafe 2 | safe >10 | **error** 2 | safe 10 | **error** 2 |
| Incr. (cas-based) | safe 7 | unsafe 2 | safe 8 | **safe** 5 | **error** 2 | **error** 2 |
| Incr./Dec. (lock-based) | safe 6 | unsafe 3 | safe >10 | **safe** >10 | safe >10 | unsafe 3 |
| Incr./Dec. (cas-based) | safe 4 | unsafe 3 | safe 6 | **safe** 8 | **error** 2 | **error** 3 |
| Tas Lock (tas-based) | safe 7 | unsafe 2 | safe 8 | **error** 2 | **error** 2 | **error** 2 |
| Ticket Lock (tas-based) | safe 8 | unsafe 3 | safe >10 | unsafe 3 | safe 5 | unsafe 3 |
| Prng (lock-based) | safe >10 | unsafe 2 | safe >10 | unsafe 2 | safe >10 | unsafe 2 |
| Prng (cas-based) | safe 5 | unsafe 3 | safe 7 | unsafe 3 | safe 6 | unsafe 3 |
| FindMax (lock-based) | safe >10 | unsafe 2 | safe >10 | **safe** >10 | safe 2 | **error** 2 |
| FindMax (cas-based) | safe 4 | unsafe 2 | safe 5 | **safe** 4 | safe 2 | **safe** 1 |
| FindMaxOpt (lock-based) | safe 7 | unsafe 2 | safe 7 | **safe** 6 | **error** 2 | **error** 2 |
| FindMaxOpt (cas-based) | safe 5 | unsafe 1 | safe 5 | unsafe 1 | **error** 2 | unsafe 1 |
| Stack (lock-based) | safe 4 | unsafe 4 | safe 4 | unsafe 4 | safe 4 | unsafe 4 |
| Stack (cas) | safe 4 | unsafe 2 | safe 4 | **safe** 6 | safe 4 | **error** 2 |

**Table 2.** Comparison of sound and unsound approaches; incorrect results in bold.

**Comparison with unsound methods.** In Section 1, we described two naïve solutions to the mixed predicate problem: uniformly using local or shared Boolean variables to represent mixed predicates, and then performing standard predicate abstraction. We de-note these approaches *mixed as local* and *mixed as shared*, respectively. Although we demonstrated theoretically in Section 1 that both methods are unsound, it is interesting to see how they perform in practice. Table 2 shows the results of applying CEGAR-based model checking to safe and unsafe versions of our benchmarks, using our sound technique, and the unsound *mixed as local* and *mixed as shared* approaches. In all cases, B-BOOM is used for model checking. For the sound technique, we show the largest thread count for which we could prove correctness of each safe benchmark, and the smallest thread count for which a bug was revealed in each unsafe benchmark. The other columns illustrate how the unsound techniques differ from this, where "error" indicates a refinement failure: it was not possible extract further predicates from spurious counterexamples. Bold entries indicate cases where the unsound approaches produce incorrect, or inconclusive results.[5] The number of cases where the unsound approaches produce false negatives, or lead to refinement failure, suggest that little confidence can be placed in these techniques, even for purposes of falsification. This justifies the more sophisticated and, crucially, sound techniques developed in this paper.

## 6   Related Work and Conclusion

There exists a large body of work on the different stages of CEGAR-based program analysis. We focus here on the abstraction stage, which is at the heart of this paper.

Predicate abstraction goes back to the foundational work by Graf and Saïdi [1]. It was first presented for *sequential* programs in a mainstream language (C) by Ball, Majumdar, Millstein, Rajamani [2] and implemented as part of the SLAM project. We have found many of the optimizations suggested by [2] to be useful in our implementation as well. Although SLAM has had great success in finding real bugs in system-level code, we are not aware of any extensions of it to concurrent programs (although this option is mentioned by the authors). We attribute this to a large part to the infeasibility, at the time, to handle realistic multi-threaded Boolean programs. We believe our own work on BOOM [5] has made progress in this direction that has made it attractive again to address concurrent predicate abstraction.

We are not aware of other work that presents solutions to the problem of "mixed predicates". Some approaches avoid it by syntactically disallowing such predicates, e.g. [19], whose authors don't discuss, however, the reasons for (or, indeed, the consequences of) doing so. In other work, "algorithmic circumstances" may make the treatment of such predicates unnecessary. The authors of [20], for example, use predicate abstraction to finitely represent the environment of a thread in multi-threaded programs. The "environment" consists of assumptions on how threads may manipulate the *shared* state of the program, irrespective of their local state. Our case of *replicated* threads, in which mixed predicates would constitute a problem, is only briefly mentioned in [20]. In [21], an approach is presented that handles *recursive* concurrent C programs. The abstract transition system of a thread (a pushdown system) is formed over predicates that are projected to the global or the local program variables and thus cannot compare

---

[5] We never expect the unsound techniques to report conclusively that a safe benchmark is unsafe: this would require demonstrating a concrete error trace in the original, safe, program.

"global against local" directly. As we have discussed, some reachability problems cannot be solved using such restricted predicates. We conjecture this problem is one of the potential causes of non-termination in the algorithm of [21].

In conclusion, we mention that building a CEGAR-based verification strategy is a tremendous effort, and our work so far can only be the beginning of such effort. We have assumed a very strict (and unrealistic) memory model that guarantees atomicity at the statement level. One can work soundly with the former assumption by pre-processing input programs so that the shared state is accessed only via word-length reads and writes, ensuring that all computation is performed using local variables. Extending our approach to weaker memory models, building on existing work in this area [22, 23], is future work. Our plans also include a more sophisticated refinement strategy, and a more detailed comparison with existing approaches that circumvent the mixed-predicates problem using other means.

## A  Boolean Broadcast Programs

Our approach to symmetry-aware predicate abstraction generates a form of concurrent Boolean program. Although languages for such programs are well-known [13, 5], the existing ones are not quite suited to our purpose. As discussed in the paper, in order to handle mixed predicates correctly, we require the facility for a Boolean program thread to read and update variables of other threads. As this is fundamental to our approach, we now present syntax and semantics for this variant of concurrent Boolean programs which we call *Boolean broadcast programs*.

```
prog ::= shared name = lit; ... ; name = lit;          lval  ::= v | [v]
         local name = lit; ... ; name = lit;
         1: stmt; ... ; k: stmt;                       expr  ::= lval | lit | compound expr

stmt ::= lval, ... , lval = expr, ... , expr           lit   ::= 0 | 1 | ⋆
       | goto pc_1, ... , pc_d  (pc_j ∈ {1, ... , k})
       | assume expr                                   name ::= any legal C variable name
```

**Fig. 1.** Syntax for Boolean broadcast programs.

Syntax for Boolean broadcast programs is specified by the grammar of Figure 1, where standard compound expressions are assumed. The language includes standard features such as shared and local variables, parallel variable assignments, nondeterministic **goto** statements and **assume** statements (which together can be used to model control flow), and the nondeterministic r-value $\star$.

The novel feature of this language is that it supports assignments of the form $[v]$ = expr, which we call *broadcasts*. When such a broadcast is executed by thread $i$, it causes local variable $v$ to be updated in each *passive* thread, i.e. in all threads *except* $i$. In this context, expr ranges over shared variables, local variables of the active thread (thread $i$), and local variables of the passive thread. The latter are distinguished by the syntax $[v]$. We refer to an expression of the form $[v]$ as a *passive expression*. We refer to such an expression as a passive l-value or passive r-value depending on which side of a broadcast it occurs.

We give semantics for parallel assignments in which all passive l-values appear after non-passive l-values. For parallel assignments where this is not the case, the semantics are defined to be the same as for any suitable rearrangement that enforces this condition. In a parallel assignment of the form:

$$u_1, \ldots, u_a, [v_1], \ldots, [v_b] = \ldots$$

we require that the $u_i$ are mutually distinct, the $v_i$ are mutually distinct, and each $v_i \in V_L$. We *do* allow the $u_i$ and $v_j$ to overlap. We also require that passive r-values only appear on the right-hand-side of broadcasts (i.e. they cannot be used in standard assignments).

We now formally define semantics for parallel assignments that may contain broadcasts. Let $\mathbb{B}$ be a Boolean broadcast program, and $n$ a positive integer. Let $V = V_S \cup V_L$ be the set of shared ($V_S$) and local ($V_L$) variable names appearing in $\mathbb{B}$. Define $\overline{V} =$

$V_S \cup \bigcup_{1 \leq i \leq n}\{l^i \mid l \in V_L\}$. This is the full set of variables in $\mathbb{B}^n$, an $n$-thread instantiation of $\mathbb{B}$. A *store* for $\mathbb{B}^n$ is a mapping $\sigma : \overline{V} \to \{0, 1\}$, assigning a Boolean value to each variable.

For an expression $\phi$, thread indices $i$ and $j$ ($1 \leq i, j \leq n$) and store $\sigma$, we write $eval(\phi, i, j, \sigma) \subseteq \{0, 1\}$ for the set of possible results which can be obtained by evaluating $\phi$ with respect to active thread $i$ and passive thread $j$ in the context of $\sigma$. Formally, for $l \in V_L$, $s \in V_S$ and $z \in \{0, 1\}$, $eval(\phi, i, j, \sigma)$ is defined for simple expressions as follows:

$$eval(z, i, j, \sigma) = \{z\} \qquad eval(\star, i, j, \sigma) = \{0, 1\}$$
$$eval(l, i, j, \sigma) = \{\sigma(l^i)\} \quad eval([l], i, j, \sigma) = \{\sigma(l^j)\} \quad eval(s, i, j, \sigma) = \{\sigma(s)\}$$

and extended to compound expressions in the obvious way.

If $\phi$ has no passive subexpressions then it is clear that, for any threads $j_1$ and $j_2$, $eval(\phi, i, j_1, \sigma) = eval(\phi, i, j_2, \sigma)$; the passive thread is irrelevant. In this case we simply write $eval(\phi, i, \sigma)$.

For a parallel assignment statement *assign*, store $\sigma$ and thread id $i$ ($1 \leq i \leq n$), we write $exec(assign, i, \sigma)$ for the set of stores that can result from active thread $i$ executing statement *assign* in the context of store $\sigma$.

Without loss of generality, we can assume that *assign* has the following form:
$$s_1, \ldots, s_a, l_1, \ldots, l_b, [m_1], \ldots, [m_c] = \phi_1, \ldots, \phi_a, \psi_1, \ldots, \psi_b, \chi_1, \ldots, \chi_c,$$
where $a, b, c \geq 0$, $s_1, \ldots, s_a \in V_S$ are mutually distinct shared variables, $l_1, \ldots, l_b \in V_L$ are mutually distinct local variables, and $m_1, \ldots, m_c \in V_L$ are also mutually distinct local variables (the $l_j$ and $m_j$ may overlap), $\phi_1, \ldots, \phi_a$ and $\psi_1, \ldots, \psi_b$ are expressions that *do not* contain passive subexpressions, and $\chi_1, \ldots, \chi_c$ are expressions that *may* contain passive subexpressions.

With this notation, we define: $exec(assign, i, \sigma) =$

$$\left\{ \begin{array}{l} \sigma[\ \ s_1\ \mapsto\ x_1\ \ , \ldots,\ s_a\ \mapsto\ x_a, \\ \phantom{\sigma[\ \ } l_1^i\ \mapsto\ y_1\ \ ,\ldots,\ l_b^i\ \mapsto\ y_b, \\ \phantom{\sigma[\ \ } m_1^1\ \mapsto\ z_{1,1}\ ,\ldots,\ m_c^1\ \mapsto\ z_{1,c},\ \ldots \\ \phantom{\sigma[\ \ } m_1^{i-1} \mapsto z_{i-1,1},\ldots, m_c^{i-1} \mapsto z_{i-1,c}, \\ \phantom{\sigma[\ \ } m_1^{i+1} \mapsto z_{i+1,1},\ldots, m_c^{i+1} \mapsto z_{i+1,c},\ \ldots \\ \phantom{\sigma[\ \ } m_1^n\ \mapsto\ z_{n,1}\ ,\ldots,\ m_c^n\ \mapsto\ z_{n,c}\ ] \end{array} \middle| \begin{array}{l} x_f \in eval(\phi_f, i, \sigma), \\ y_g \in eval(\psi_g, i, \sigma), \\ z_h^j \in eval(\chi_h, i, j, \sigma), \\ 1 \leq f \leq a, 1 \leq g \leq b, \\ 1 \leq h \leq c, 1 \leq j \leq n, i \neq j \end{array} \right\}$$

Thus each store in $exec(assign, i, \sigma)$ is derived from $\sigma$ by setting each variable $s_f$ and $l_g^i$ to a value of $eval(\phi_f, i, \sigma)$ and $eval(\psi_g, i, \sigma)$ respectively ($1 \leq f \leq a, 1 \leq g \leq b$), and, for each thread $j$ distinct from $i$, setting variable $m_h^j$ to a value of $eval(\chi_h, i, j, \sigma)$ ($1 \leq h \leq c$).

This definition makes precise the meaning of the $[.]$ notation used in the paper. The definitions of $exec$ and $eval$ can be used to define the transition system associated with a Boolean broadcast program in the standard way.

# B  Proving $\mathbb{P}''$ Correct Requires Mixed Predicates

Recall program $\mathbb{P}''$ from Section 1, which, as we show, we cannot prove correct if executed by 1 thread.

$\mathbb{P}''$:
```
0: shared int r = 0;
   shared int s = 0;
   local  int l = 0;
1: ++r;
2: if (r == 1) then
3:    f();
```

$f()$:
```
4: ++s, ++l;
5: assert s == l;
6: goto 4;
```

Let $E = E^{r,s} \cup E^l$ for **disjoint** sets $E^{r,s}$ and $E^l$ of predicates over $\{r,s\}$ and $l$, resp.; in particular, no predicate refers to both $s$ and $l$. Suppose $I$ is an invariant of $\mathbb{P}''$ expressible over predicates in $E$ such that $I \Rightarrow (s{=}{=}l)$ is valid. Since every state satisfying $r{=}{=}1$, $s{=}{=}l$ is reachable in $\mathbb{P}''$, $(r{=}{=}1) \wedge (s{=}{=}l) \Rightarrow I$ is valid. Therefore, for infinitely many $c \in \mathbb{N}$, the assignment $r{=}1$, $s{=}l{=}c$ satisfies $I$, written $(r,s,l){=}(1,c,c) \models I$.

Let now $\{I_1, \ldots, I_w\}$ be the cubes in the DNF representation of $I$. Since this set is finite, there exist two distinct values $a,b$ and some $i \in \{1,\ldots,w\}$ such that both $(r,s,l){=}(1,a,a) \models I_i$ and $(r,s,l){=}(1,b,b) \models I_i$. We split cube $I_i$ into the sub-cubes $I_i^{r,s}$ and $I_i^l$ that contain the predicates over $\{r,s\}$ and those over $l$, resp.: $I_i = I_i^{r,s} \wedge I_i^l$. From $(r,s,l){=}(1,a,a) \models I_i$ we conclude $(r,s){=}(1,a) \models I_i^{r,s}$ ($l$ does not occur in $I_i^{r,s}$). Similarly, from $(r,s,l){=}(1,b,b) \models I_i$ we conclude $l{=}b \models I_i^l$. Hence, $(r,s,l){=}(1,a,b) \models I_i$, hence $(r,s,l){=}(1,a,b) \models I$, which contradicts the validity of $I \Rightarrow (s{=}{=}l)$ since $a \neq b$. $\square$